\documentclass[sigconf,screen]{acmart}
\usepackage{booktabs}
\usepackage{listings}
\usepackage{dirtree}
\usepackage{forest}
\usepackage{caption}
\usepackage{subcaption}

\copyrightyear{2022}
\acmYear{2022}
\setcopyright{acmlicensed}\acmConference[RTNS '22]{Proceedings of the 30th International Conference on Real-Time Networks and Systems}{June 7--8, 2022}{Paris, France}
\acmBooktitle{Proceedings of the 30th International Conference on Real-Time Networks and Systems (RTNS '22), June 7--8, 2022, Paris, France}
\acmPrice{15.00}
\acmDOI{10.1145/3534879.3534888}
\acmISBN{978-1-4503-9650-9/22/06}



\definecolor{folderbg}{RGB}{124,166,198}
\definecolor{folderborder}{RGB}{110,144,169}

\def\Size{4pt}
\tikzset{
  folder/.pic={
    \filldraw[draw=folderborder,top color=folderbg!50,bottom color=folderbg]
      (-1.05*\Size,0.2\Size+5pt) rectangle ++(.75*\Size,-0.2\Size-5pt);  
    \filldraw[draw=folderborder,top color=folderbg!50,bottom color=folderbg]
      (-1.15*\Size,-\Size) rectangle (1.15*\Size,\Size);
  }
}

\newcommand{\pgrph}[2]{\noindent\emph{\textbf{#1~}}{#2}}
\newcommand{\yes}[0]{{\color[HTML]{009901} \boldmath\textbf{$\checkmark$}}}
\newcommand{\nop}[0]{{\color[HTML]{FE0000} \boldmath\textbf{$\times$}}}
\newcommand{\meh}[0]{{\color[HTML]{F56B00} \boldmath\textbf{-}}}
\newcommand{\orange}[1]{{\color[HTML]{F56B00} #1}}

\newcommand{\newtext}[1]{#1}

\title{RT-Bench: an Extensible Benchmark Framework for the Analysis and Management of Real-Time Applications}
\begin{document}

\lstset{numbers=left,breaklines=true,frame=single}

\author{Mattia Nicolella}
\affiliation{\institution{\small Boston University\country{USA}}}
\email{mnico@bu.edu}

\author{Shahin Roozkhosh}
\affiliation{\institution{\small Boston University\country{USA}}}
\email{shahin@bu.edu}

\author{Denis Hoornaert}
\affiliation{\institution{\small TU M\"unchen\country{ Germany}}}
\email{denis.hoornaert@tum.de}

\author{Andrea Bastoni}
\affiliation{\institution{\small TU M\"unchen\country{ Germany}}}
\email{andrea.bastoni@tum.de}

\author{Renato Mancuso}
\affiliation{\institution{\small Boston University\country{USA}}}
\email{rmancuso@bu.edu}
\begin{abstract}
Benchmarking is crucial for testing and validating any system, including---and perhaps especially---real-time systems.
Typical real-time applications adhere to well-understood abstractions: they exhibit a periodic behavior, operate on a well-defined working set, and strive for stable response time, avoiding non-predicable factors such as page faults.
Unfortunately, available benchmark suites fail to reflect key characteristics of real-time applications. Practitioners and researchers must resort to either benchmark heavily approximated real-time environments or re-engineer available benchmarks to add---if possible---the sought-after features.
Additionally, the measuring and logging capabilities provided by most benchmark suites are not tailored ``out-of-the-box'' to real-time environments, and changing basic parameters such as the scheduling policy often becomes a tiring and error-prone exercise.

In this paper, we present RT-bench, an open-source framework adding standard real-time features to virtually \emph{any} existing benchmark. Furthermore, RT-bench provides an easy-to-use, unified command-line interface to customize key aspects of the real-time execution of a set of benchmarks.
Our framework is guided by four main criteria: 1) cohesive interface, 2) support for periodic application behavior and deadline semantics, 3) controllable memory footprint, and 4) extensibility and portability. We have integrated within the framework applications from the widely used SD-VBS and IsolBench suites. We showcase a set of use-cases that are representative of typical real-time system evaluation scenarios, and that can be easily conducted via RT-Bench.
\end{abstract}

\maketitle
\section{Introduction}
\label{sec:intro}
In light of the current ever-growing dependence on automated control systems and their increasing complexity, with numerous components consolidated on a chip, safety and determinism certification challenges grow exponentially. At the early stages of development, simulation tools are essential to inspect new designs. However, simulation of the entire system works under the assumption that the employed models accurately represent the actual behavior of the hardware. Unfortunately, complex modern hardware often deviates from textbook models in unpredictable ways as they only reveal partial information on the actual state of the system~\cite{tarapore2021observing}.
Real-time researchers are therefore forced to analyze the system's deployment on the final physical platforms. Benchmark suites are, therefore, frequently used to bridge the gap between simulated and real behavior and empirically assess the ability to deliver real-time performance.

In response to the challenges outlined above, the real-time community has adapted to use miscellaneous sets of techniques to test various system components both before and after integration. The latter is often more demanding as it requires to reason about the interplay between concurrent software components. To that end, numerous benchmarks have been used to test a multitude of characteristics. Unfortunately, however, the lack of standardized benchmarks and general consensus in the testing environment has led to severe fragmentation in testing methodology and poor comparability of results. To make things worse, specific base-platform dependencies and ordeals in porting benchmarks to run on specific hardware are often overlooked challenges that can adversely affect the ability to adopt a given set of benchmarks.

When looking at a real-time platform, metrics such as responsiveness, predictability, and the impact of parallelism are of immediate interest. It is also customary to analyze real-time systems under \emph{typical} load as well as under \emph{stress}.
Therefore, a good practice is to use a mixture of both synthetic and realistic benchmarks to construct an informed assessment of how the system reacts to different workload configurations. Despite the abundance of benchmarks of both types, there is a lack of out-of-the-box applicability to real-time systems.

This paper aims to propose a standard framework, namely RT-Bench, that offers several features designed to meet the needs of researchers and practitioners who are interested in studying the real-time behavior of their systems.
Beyond providing an initial set of \newtext{well-known and publicly available} benchmarks\footnote{\newtext{San-Diego Vision Benchmark suite~\cite{SD-VBS}}} that \newtext{have been adapted to RT-Bench}, our goal is to allow the integration of additional applications with as few adaptations as possible in a modular and extensible fashion that emphasizes the importance of good documentation and code reuse.
We identify three main categories of functional features: (1) unified launch, control, and reporting interface; (2) adherence to real-time abstractions; (3) cross-platform portability; and (4) automated analysis use-cases.

\pgrph{Unified Interface:} RT-Bench provides system designers with the direct ability to control key parameters of benchmark deployment, such as workload composition, scheduling policy and priorities, pinning of applications to CPUs, and enforcement of memory allocation limits, to name a few. At the same time, the framework provides a uniform performance reporting infrastructure that already includes real-time oriented metrics such as job arrival time, deadline, response time, and usage of system resources via performance counters. Lastly, RT-Bench includes a set of automated analysis use-cases aimed at producing key complex metrics in the platform of reference, such as observed WCET, the impact of contention-induced temporal interference, and input-dependent working set size (WSS).

\pgrph{Real-time Abstractions:} RT-Bench proposes an infrastructure to transform any monolithic benchmark into a recurrent one with the goal of adhering to the real-time periodic task model. In doing so, it also offers a generic methodology to factor-out typical initialization and tear-down overheads in the acquired measurements. At the same time, real-time applications are often assumed to be deadline-constrained. For this purpose, RT-Bench implements deadline detection and enforcement semantics via job skipping. Moreover, real-time applications are assumed to have well-behaved memory allocation patterns and statically known WSS. RT-Bench offers a generic technique to enforce such semantics even when the original benchmarks make use of dynamic memory allocation (e.g., via \texttt{malloc} and \texttt{free}) and without requiring code refactoring.

\pgrph{Cross-platform Portability:} RT-Bench only uses APIs from the POSIX standard to allow deployment on an extensive range of Operating Systems (OS) and bare-metal software stacks (e.g. Newlib~\cite{newlib}). We decouple it from any system-specific limitations through user application-level implementation.
The only exception is user-space interaction with platform-specific cycle counters for which support is provided in all the leading architectures.

We test our framework by adapting to the state-of-the-art benchmark suits as we explore their characteristics in the remainder of this paper. A proof-of-concept integration of IsolBench~\cite{IsolBench} and San Diego Vision Benchmark Suite (SD-VBS)~\cite{SD-VBS} into the proposed framework is provided. The open-source RT-Bench implementation is available at \url{https://gitlab.com/rt-bench/rt-bench}.
\section{Related Work}
\label{sec:related}

With the ever-growing explosion in the complexity of embedded computing platforms, performance characterization and prediction have become increasingly more challenging. Moreover, to reach a conclusive assessment regarding the temporal properties of a system, it is crucial to test the system's behavior under different workloads. The real-time community has adopted a number of strategies to obtain indicators of the system behavior through benchmarking.

\begin{table*}[!htp]
    \caption{Benchmark suites comparison (ALL = Synthetic, Pragmatic, and Full-Scale)}
    \vspace{-0.3cm}
    \label{tab:feat-comp}
    \begin{tabular}{@{}lcccccccccccc@{}}
        \toprule
        \begin{tabular}[c]{@{}c@{}}Benchmark \\ suite\end{tabular} &
        Type &
        \begin{tabular}[c]{@{}c@{}}Periodic\\ exec.\end{tabular} &
        \begin{tabular}[c]{@{}c@{}}Cross \\ Platform\end{tabular} &
        \begin{tabular}[c]{@{}c@{}} Unified \\ Interface\end{tabular} &
        \begin{tabular}[c]{@{}c@{}} Dead. \\ status\end{tabular} &
        \begin{tabular}[c]{@{}c@{}} Exec. \\ Time\end{tabular} &
        \begin{tabular}[c]{@{}c@{}} Utili-\\zation \end{tabular} &
        Density &
        \begin{tabular}[c]{@{}c@{}} \newtext{Profiled}\\\newtext{WCET} \end{tabular} &
        \begin{tabular}[c]{@{}c@{}} \newtext{Profiled}\\\newtext{WSS} \end{tabular} &
        \begin{tabular}[c]{@{}c@{}} Perf.\\counters \end{tabular} &
        \begin{tabular}[c]{@{}c@{}} Memory \\ Profiler\end{tabular} \\
        \midrule
        TACLeBench~\cite{taclebench}             & ALL   & \meh & \meh & \nop & \nop & \nop & \nop & \nop &\nop & \nop & \nop & \nop \\
        SD-VBS~\cite{SD-VBS}                     & PB    & \nop & \yes & \nop & \nop & \yes & \nop & \nop &\nop & \nop & \nop & \nop \\
        Mälardalen~\cite{malardalen}             & PB    & \nop & \yes & \nop & \nop & \nop & \nop & \nop &\yes & \nop & \nop & \nop \\
        SPLASH~\cite{splash2x, splash3}          & PB    & \nop & \orange{CUDA} & \nop & \nop & \yes & \nop & \nop &\nop & \yes & \nop & \nop \\
        Rodinia~\cite{rodinia}                   & PB    & \nop & \yes & \nop & \nop & \meh & \nop & \nop &\nop & \nop & \nop & \nop \\
        PARSEC~\cite{parsec}                     & PB    & \nop & \yes & \nop & \nop & \yes & \nop & \nop &\nop & \nop & \nop & \nop \\
        IsolBench~\cite{IsolBench}               & SB    & \yes & \yes & \nop & \nop & \yes & \nop & \nop &\nop & \nop & \nop & \nop \\
        EEBMC~\cite{EEMBC}              & \newtext{PB}  & \nop & \yes & \nop & \nop & \yes & \nop & \nop &\nop & \nop & \nop & \nop \\
        MiBench~\cite{mibench}          & \newtext{PB}  & \nop & \yes & \nop & \nop & \nop & \nop & \nop &\nop & \nop & \nop & \nop \\
        PapaBench~\cite{papabench}      & FS    & \yes & \orange{AVR} & \nop  & \nop & \nop & \nop & \nop &\nop & \nop & \nop & \nop \\
        RT-Tests \cite{rt-test}         & PB/SB & \orange{some} &\yes & \nop & \nop & \yes & \nop & \nop & \orange{some} & \nop & \nop & \nop \\
        RTEval \cite{rteval}            & SB    & \nop & \yes & \nop & \nop & \yes & \nop & \nop & \yes & \nop & \nop & \nop \\
        RT-bench                        & PB/SB & \yes &\orange{ARM/x86} & \yes & \yes & \yes & \yes & \yes & \orange{script} & \orange{script} & \orange{Perf} & \orange{Aarch64} \\
        \bottomrule
    \end{tabular}
    \vspace{-0.3cm}
\end{table*}

This section provides a comparative survey of popular benchmark suites used in the real-time community. The survey is summarized in Table~\ref{tab:feat-comp}. These suites can be categorized into three groups:
\begin{itemize}
    \item \textbf{Synthetic Benchmarks (SB)} that will stress a particular element or aspect of the system under analysis.
    \item \textbf{Pragmatic Benchmarks (PB)} batch processing tasks mimicking realistic workload such as image processing, signal processing, physics simulation, and matrix multiplication.
    \item \textbf{Full-Scale (FS)} real-time applications containing a mixture of hard/soft/non-realtime jobs with both periodic and non-periodic tasks to be executed on embedded systems for full-system concrete timing verification.
\end{itemize}
Popular \textbf{Synthetic} benchmarks include IsolBench~\cite{IsolBench} (used in~\cite{9113094,ghaemi_et_al:LIPIcs.ECRTS.2021.4,hoornaert_et_al:LIPIcs.ECRTS.2021.2}),
\newtext{the RT-Test \cite{rt-test} suite and the RTEval \cite{rteval} benchmark.}
IsolBench is a collection of memory workloads used to analyze the memory bandwidth and latency. It supports periodic execution, but it does not have a comprehensive interface logging data on each period.
\newtext{The RT-Test suite is a set of benchmarks to profile the responsiveness of the Linux kernel. The RTEeval benchmark relies on RT-Test to measure the performance of the Linux kernel under specific workloads.}

The most accustomed \textbf{Pragmatic} benchmark suites include: TACLeBench~\cite{taclebench} (used in~\cite{tessler2020bringing,kadar2021safety,markovic2020improving}), San Diego Vision Benchmarks~\cite{SD-VBS} (used in~\cite{bakita2021simultaneous,9113094,roozkhosh2020potential}), Mälardalen~\cite{malardalen} (used in~\cite{bellec2020attack,markovic2020cache,markovi_et_al:LIPIcs:2020:12368}), several versions of SPLASH~\cite{splash2,splash2x,splash3} (used in~\cite{gifford2021dna,wu2021hardware,kaushik2021systematic}), EEMBC~\cite{EEMBC} (used in~\cite{hassan:LIPIcs:2020:12379,hassan_et_al:LIPIcs:2020:12386}), and MiBench~\cite{mibench} (used in~\cite{islam2020scheduling,cui2021fault}).
TACLeBench has been designed with portability in mind for most of the benchmarks, as they target WCET analysis. This suite regroups other synthetic benchmarks such as Papabench~\cite{papabench}. Hence,  TACLeBench lacks a homogeneous interface.
SD-VBS performs general image processing and vision-related jobs, and it aims at offering maximum portability. However, to be used as embedded applications, the benchmarks must be adapted for periodic execution. Dynamic memory allocation is also widely used, which further hinders their temporal determinism.
The Mälardalen benchmarks share many components with TACLeBench. However, they also suffer from some of the same shortcomings. The Mälardalen benchmarks are not designed for periodic execution. Instead, they are mainly designed to be good targets for static WCET analysis.
\newtext{EEBMC is a selection of benchmarks specifically targeting embedded devices of different types, ranging from mobile to automotive systems. MiBench is similar to EEMBC, but was created to address its shortcomings.}
Finally, the SPLASH benchmark suite is a collection of benchmarks tailored to parallel execution and WSS analysis from Pragmatic models. On a similar flavor, suites like the PARSEC~\cite{parsec} (used in~\cite{gifford2021dna}) and Rodinia~\cite{rodinia} suites (used in~\cite{olmedo2020dissecting,bateni2020co}) represent an interesting alternative as they specifically target parallel execution, with Rodinia even offering support for GPUs and heterogeneous systems.

\textbf{Full-scale} real-time applications mostly come from the WATERS industrial challenge \cite{hamann2017waters}, formerly called Formal Methods for Timing Verification (FMTV). Full-scale applications are, for the most part, periodic, but extending them is complex, and they might not have broad multi-platform support. An example of a full-scale real-time application is the Papabench~\cite{papabench} benchmark, which encapsulates all the main components of a control system for UAVs.

\autoref{tab:feat-comp} summarizes the essential characteristics \newtext{(columns)} of the surveyed benchmarks \newtext{(rows)}. \newtext{The reported characteristics include (1) the \emph{Type} of benchmark the suite offers according to the aforementioned categories, (2) the capability to be executed in a \emph{Periodic} fashion, (3) the provided cross-platform support, (4) whether it provides a unified interface with other suites, and (5) the metrics natively reported by the suites. For the latter, this includes, from left to right, whether the deadline has been met, the execution time, the processor utilization\footnote{Computed as measured execution time over the period.}, the density\footnote{Computed as measured execution time over the relative deadline.}, the empirically observed WCET, the ability to collect and report end-to-end hardware events obtained though performance counters (e.g., cache accesses), and the ability to monitor the trend of observed hardware events throughout the execution. Note that categories for which no clear-cut answer exists are marked in orange. This is the case for the platforms supported, and the test provided by RT-Bench noted as \emph{\orange{script}}, meaning that they rely on high-level tools.} \newtext{The table} highlights the necessity of a framework that is specifically designed for the analysis of real-time systems. Indeed, the core philosophy of the proposed RT-Bench framework is to provide an infrastructure to build a reference set of real-time benchmarks with standard functionalities. As a first step in this direction, RT-Bench already offers key analysis tools such as execution-time distribution analysis, WSS examination, and sensitivity to interference. Moreover, with RT-Bench, existing benchmarks can be integrated to execute periodically and to exhibit \newtext{controlled} memory \newtext{footprint} with minimum re-engineering effort.

\section{Design Goals and Overview}
    \label{sec:design}
    As presented in Section~\ref{sec:related}, a real-time analysis should ideally be conducted using a large set of benchmarks with different characteristics to provide a comprehensive understanding of the (real-time) performance of the system under analysis. With that respect, the objective of the proposed RT-Bench framework is three-fold.
    
    \vspace*{0.1cm}
    \pgrph{Common and cohesive interfaces.}{The use of benchmark suites is widespread in the real-time community, and it is not rare for multiple suites to be jointly used in a given study. These suites are, in most cases, contributions from distinct individuals having particular focuses, ranging from CPU- or memory-bound to CPU- or memory-intensive applications. Unfortunately, while this diversity is a strength, it entails a fragmentation of the parameters available (e.g., assigned processing units, scheduling policy), the metrics reported (e.g., response time, working set size), and the overall user experience. RT-Bench aims at bridging this gap by homogenizing the available features and the reports generated for any benchmark by offering a unified and coherent interface.}
    
    \vspace*{0.1cm}
    \pgrph{Adherence to Real-Time System Abstractions}{We aim to incorporate, within the proposed RT-Bench framework, a set of features that are in line with the typical models and assumptions used for research, analysis, and testing of real-time systems. We consider this objective of paramount importance and a clear distinctive factor compared to the surveyed suites. RT-Bench is deliberately designed from the ground up to transform any one-shot benchmark into a periodic application with deadline enforcement and job-skipping semantics, with compartmentalized one-time initialization and tear-down routines, so to obtain precise measurements. In addition, any benchmark integrated within RT-Bench natively features options to control allocation on a specific set of cores; to be assigned a scheduling policy, and to limit and pre-allocate memory. These features effectively align the behavior of RT-Bench applications with a critical mass of assumptions and abstraction that are customary when analyzing real-time systems.}
 
    \vspace*{0.1cm}
    \pgrph{Extensibility, portability, and usability.}{We carefully designed the proposed framework, RT-Bench, to be easily extensible, portable, and practical. We deliberately implemented the RT-Bench core features to operate in user space so as to decouple our framework from any system-specific constraints. We do so by leveraging widespread POSIX system-level interfaces. Doing so enables RT-Bench benchmarks to be deployed on a wide range of OS's and bare-metal software stacks (e.g. Newlib). There are only two exceptions to this rule which correspond to two advanced features provided by the framework. The first is the ability to gather timing statistics directly from architecture-specific performance counters. In this case, assembly functions to support x86, Aaarch32, and Aaarch64 systems have already been included. Second, the possibility to also gather statistics from performance counters relies on the Perf~\cite{perf} infrastructure, which is available by default in typical Linux kernels. To enhance usability, we also provide a complete set of automated build scripts. Likewise, we include a large set of on-the-fly/post-processing scripts.}
    
    \vspace*{0.1cm}
    \newtext{The} RT-Bench \newtext{framework} comprises three specific components: (1) the RT-Benchmark Generator (mandatory), (2) Utils, and (3) Measurements Processing tools. Only RT-Benchmark Generator is mandatory, while the rest are optional. These modules, as mentioned above, are described in the remainder of this section. We further explore their purposes and interaction.
    
    \subsection{RT-Benchmark Generator}
        \label{sec:rtbench_core}
        
        RT-Bench is designed to be extended with additional third-party benchmarks. Towards this goal, any ported benchmark shall follow the same interface and shall support the same real-time features mentioned in this section.
        
        \begin{figure}[t!]
            \centering
            \includegraphics[scale=0.175]{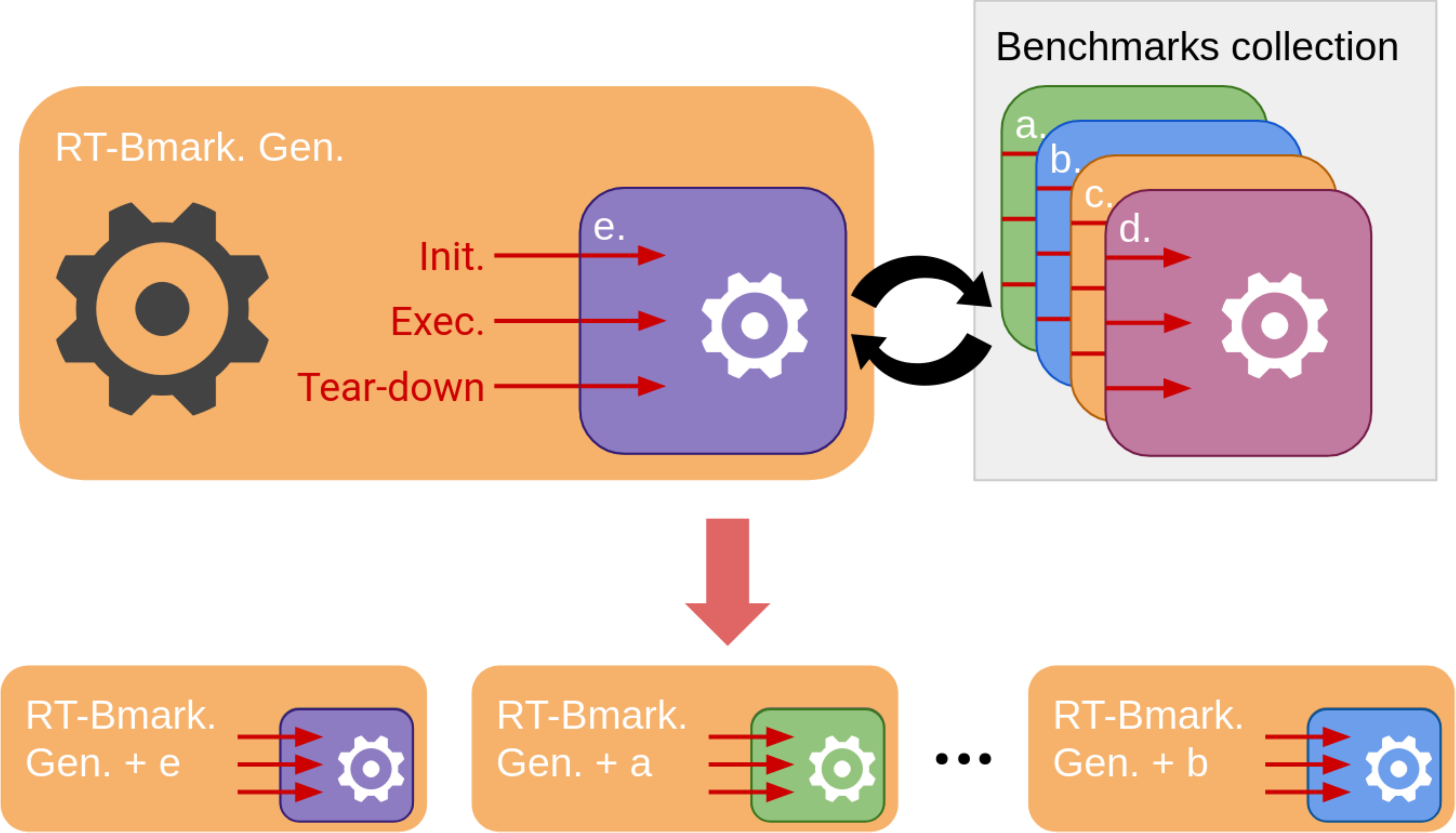}
            \caption{The RT-Benchmark Generator can simply be extended with any benchmark from a collection (e.g., \emph{a}, \emph{b}, \emph{c}, \emph{d}, and \emph{e}) as long as they feature the three harnessing points constituting the interface. It yields RT-Benchmarks version of the adapted third-party benchmarks.}
            \label{fig:rt-benchmark-generator}
        \end{figure}
        
        The conversion to the RT-Bench format is near-transparent, as it only requires the benchmark to be slightly altered to comply with the proposed interface. The interface consists of three functions acting as harness points: (1) \emph{Initialization}, (2) \emph{Execution}, and (3) \emph{Tear-down}.
        \newtext{These functions---that must be implemented for a benchmark to be integrated within RT-Bench---}are respectively in charge of (1) initializing shared resources such as memory, file descriptors, shared data objects, and the like, (2) executing the main application logic/algorithm, and (3) freeing any of the resources used.
        Their exact utilization is, from the standpoint of the benchmark, opaquely driven by the RT-Benchmark Generator (see Section~\ref{subsec:benchmarking-mechanism}), effectively decoupling real-time features from the design of the application at hand.
        Further details regarding the implementation are provided in Section~\ref{subsec:harnessing_functions_and_extension}.
        As illustrated in Figure~\ref{fig:rt-benchmark-generator}, once the benchmark to be ported is structured following the interface outlined above, the build scripts automate the creation of stand-alone executables that include all the top-level features implemented by the RT-Benchmark Generator. Therefore, encapsulating the desired benchmark within RT-Bench transparently and effortlessly grants it a uniform set of features and a coherent launch interface.
        
        \emph{Periodic execution} is an essential feature of the framework, as it ensures a periodic execution of the benchmark's main algorithm for a specified amount of iterations---potentially infinitely many. The periodic executions are coherent with the user-specified deadline, meaning if the task does not complete, its successor is not released, and the deadline miss is reported---i.e., RT-Bench applications adhere to the job skipping~\cite{job-skipping} approach to handle any detected overload conditions.
        
        \emph{Core and scheduling policy selection} is provided to perform partitioned and clustered multi-core scheduling through pinning to a specific set of cores. 
        The range of execution units and policies available depends on the considered platform. On typical Linux kernels, RT-Bench allows the selection of scheduling policies such as \texttt{SCHED\_OTHER}, \texttt{SCHED\_FIFO}, \texttt{SCHED\_RR}, and \texttt{SCHED\_DEADLINE} and corresponding parameters.

        \begin{table*}[]
            \caption{List of metrics logged by RT-Bench and their units.}
            \vspace{-0.3cm}
            \label{tab:logged_data}
            \begin{tabular}{llll}
                \toprule
                Metric                    & Description                                       & Formula                                          & Unit                    \\
                \midrule
                \texttt{period\_start}    & Period start timestamp                            &                                                  & ns and CPU clock-cycles \\
                \texttt{period\_end}      & Period end timestamp                              &                                                  & ns and CPU clock-cycles \\
                \texttt{job\_end}         & Job end timestamp                                 &                                                  & ns and CPU clock-cycles \\
                \texttt{deadline}         & Absolute deadline timestamp                       &                                                  & ns and CPU clock-cycles \\
                \texttt{deadline\_met}    & Status of the deadline. 1 if met, 0 otherwise.    &                                                  & Boolean                 \\
                \texttt{job\_elapsed}     & Absolute job response time                      & $job\_end-period\_start$                         & ns and CPU clock-cycles \\
                \texttt{job\_utilization} & Job utilization                                   & $\frac{job\_elapsed}{period\_end-period\_start}$ & Ratio                   \\
                \texttt{job\_density}     & Job density                                       & $\frac{job\_elapsed}{deadline-period\_start}$    & Ratio                   \\
                \texttt{l1\_references}   & L1 References (PMC)                               &                                                  & Absolute number         \\
                \texttt{l1\_refills}      & L1 Refills (PMC)                                  &                                                  & Absolute number         \\
                \texttt{l2\_references}   & L2 References (PMC)                               &                                                  & Absolute number         \\
                \texttt{l2\_refills}      & L2 Refills (PMC)                                  &                                                  & Absolute number         \\
                \texttt{inst\_retired}    & Instructions retired (PMC)                        &                                                  & Absolute number         \\
                \toprule
            \end{tabular}
            \vspace{-0.3cm}
        \end{table*}
        
        A \emph{deterministic memory layout} is important for real-time applications. Indeed, one often wants to study the memory footprint (or working set size) of the benchmark under analysis and to avoid the overhead of page faults and swapping.
        \newtext{While the RT-Bench framework cannot provide a deterministic memory allocation for applications using dynamic memory (e.g., via \texttt{malloc} and \texttt{free}), it instead enforces a \emph{deterministic memory layout} with two-fold semantics to control memory allocation. When enabled, the user must specify a maximum amount of heap memory to be pre-allocated. All the specified memory is physically allocated (faulted-in) and locked (i.e., made non-swappable) at initialization. Additionally, a watchdog routine is installed to (1) monitor the actual benchmark's footprint at each memory allocation, (2) disable the creation of additional virtual memory regions, and (3) enforce a strict size limit on the heap region, terminating any application exceeding it.}
        
        Finally, RT-Bench offers a common \emph{reporting (output) interface} to export the data collected throughout the execution. The metrics listed in Table~\ref{tab:logged_data} can be reported in four verbosity levels: (1) error-only; (2) full logging in a CSV file format; (3) full logging on the standard output; and (4) full logging on the standard output in a human-readable format.
        
        Encapsulating the target benchmark within RT-bench means that any ported benchmark benefits from all the aforementioned features. Moreover, they natively display the command-line options to set any of the required parameters. This is ultimately what allows all the applications to share a \textbf{standard and coherent launch interface} throughout the RT-Benchmark collection.
        The entirety of the discussed features (and command-line options) are further discussed in Section~\ref{subsec:using_rtbench} and exhaustively listed in the project documentation. 
        
    \subsection{Measurements Processing}
        \label{sec:high-level-data-processing}
        
        Alongside the mandatory core module, a.k.a. the RT-Benchmark Generator, the framework also includes a series of optional high-level scripts built on top of the generator.
        The provided scripts are written with high-abstraction-level languages such as \emph{python} and \emph{bash}. They aim to provide a well-rounded user experience in at least four ways:
        (1) they automatically perform common tasks such as \newtext{empirically determining} a benchmark's WSS, WCET, and ACET; (2) they ease the launch of interfering tasks, both memory- and CPU-intensive on both the same or other available CPUs; (3) they perform system-dependent preparation tasks such as migrating and pinning on selected CPUs to limit undesired interference; and (4) they generate plots of the obtained results using plotting libraries.
        
        The script set is a prime example of tools exploiting the RT-Bench standard interfaces, setting the benchmark parameters following the standard command options, and extracting the measurements by parsing the reporting format.
        
    \subsection{Utils}
        \label{sec:utils}
        The RT-Bench framework also comes with project maintenance and deployment tools, further improving portability and usability.
        
        The framework provides a fully automated build system to generate RT-Bench benchmarks for each supported suite. It enables the building and management of suites individually and globally.
        Cross-platform compiling is supported for \texttt{ARM} and \texttt{x86\_64}.
        
        Additionally, complete documentation regarding the framework's RT-Benchmark generator is provided. This documentation is available in both HTML and \LaTeX~ locally and on the framework's repository. It is generated by Doxygen~\cite{doxygen} and already available online\footnote{See \url{https://bastoni.gitlab.io/rt-bench/}.}.
\section{Implementation}
    \label{sec:implmentation}
    This section presents the main technical details behind the implementation of our RT-Bench. This section focuses on the RT-Benchmark Generator, its mechanisms, and how it must be used to port a \emph{generic} monolithic benchmark. Later in the section, the emphasis is put on the optional side tools offered with the framework to streamline common real-time oriented tests.
    
    As stated in earlier sections, RT-Bench has been designed with extensibility and portability in mind. This has led to some determining implementation decisions. The implementation presented in-depth in this section and evaluated in Section~\ref{sec:eval} assumes that Linux is the OS of reference. Even though most of the features only depend on POSIX, other features such as the available real-time scheduling policies are inherently dependent on the OS. Selecting Linux provides us with a sound selection of policies (e.g., \texttt{SCHED\_FIFO}, \texttt{SCHED\_DEADLINE}) and a simple interface to configure their parameters (i.e., \texttt{SYS\_sched\_*} syscalls).
    
    \subsection{Core Mechanism}
        \label{subsec:benchmarking-mechanism}
        \begin{figure}[t]
            \includegraphics[scale=0.49]{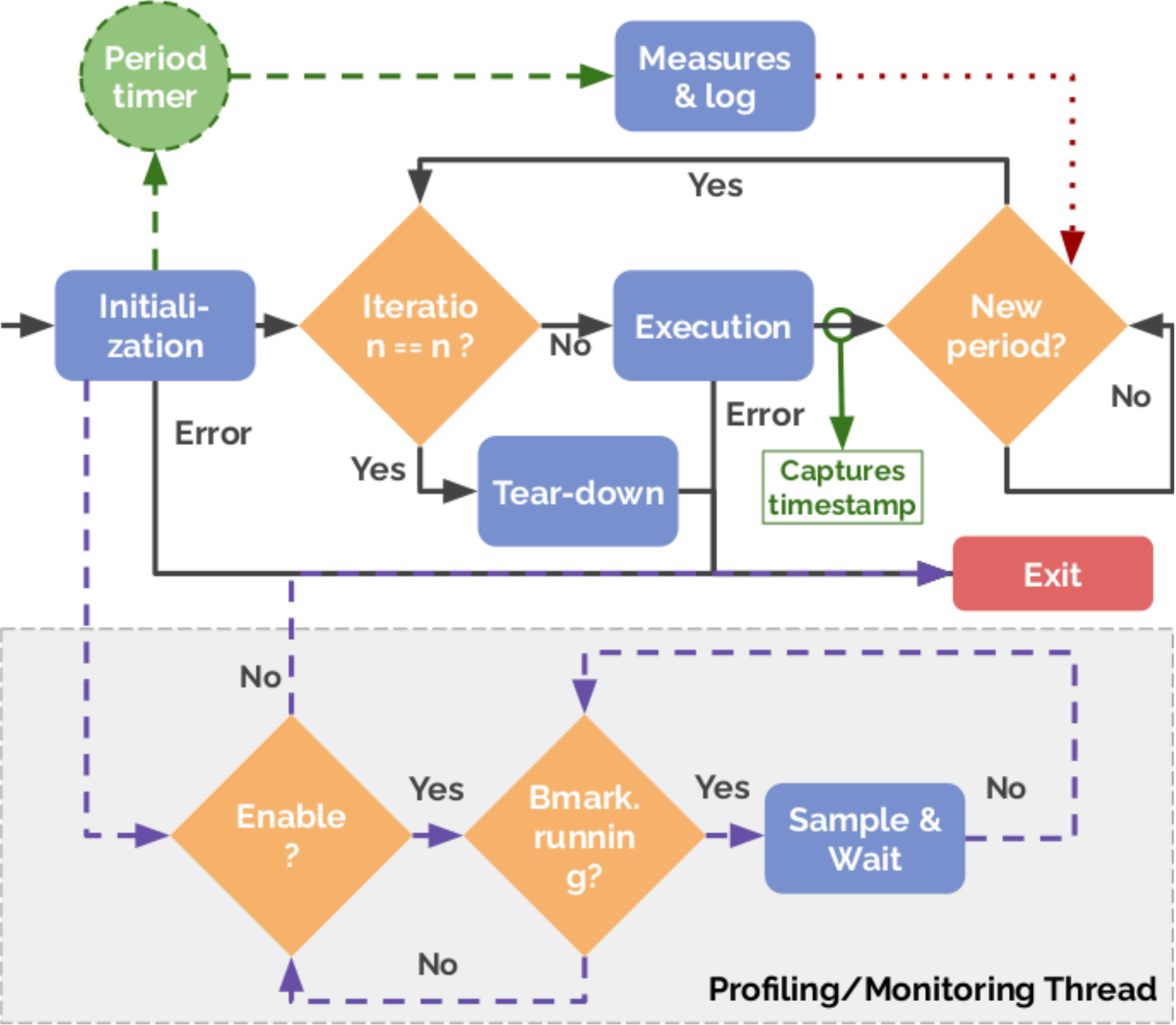}
            \caption{Flowchart of the mechanism used by RT-Bench.}
            \label{fig:rt-bench-logic-flowchart}
        \end{figure}
        
        At the heart of any of the benchmarks generated using RT-Bench lays the logic and mechanism in charge of enabling the desired features listed in Section~\ref{sec:design}.
        RT-Benchmark generator is responsible for invoking the entry points (to be implemented as part of the porting of a new benchmark) at adequate moments. This enables RT-Bench to provide the features described in Section~\ref{sec:design} to any compliant benchmark it is attached to.
        
        A flow-graph representation of said logic is shown in Figure~\ref{fig:rt-bench-logic-flowchart}. \newtext{The core logic is executed as a single-threaded process.} The first step (or entry point) in the RT-Bench logic broadly consists of the \textbf{initialization} of the benchmarking environment. In addition to calling the associated benchmark's \emph{Initialization} harnessing function, this initialization phase sets up every feature provided by RT-Bench using the user-specified inputs or the default ones. For instance, this includes the period, the deadline, or the amount of iterations. Noticeably, a timer-triggered (using a high-resolution timer) signal with the specified periodicity is set up.
        In Fig.~\ref{fig:rt-bench-logic-flowchart}, the timer is attached to the main thread, and its transitions are dashed and colored in green. At any point, if an error arises, a message is provided in output, and the benchmark is terminated (see \texttt{Exit} in Figure~\ref{fig:rt-bench-logic-flowchart}).
        Thereafter, the benchmark is ready to enter its periodic \textbf{execution} phase. The amount of iterations specified by the user ($n$) is enforced. There are two possible outcomes: the desired amount of iterations has been reached, or a few iterations remain to be performed. In the latter case, the benchmark is executed by calling the \textbf{Execution} harnessing function. Upon completion of the benchmark's workload execution, the process is blocked until a new period starts. In such case, the process loops back to the iteration comparison.
        \newtext{Only iterations in which a job was started are considered in the comparison to make sure that $n$ jobs are executed.}
        Once all execution iterations have been performed, the benchmark can terminate gracefully by entering its \emph{tear-down} phase, effectively calling the \textbf{Tear-down} harnessing functions.
        
        Note that the \emph{Initialization} and \emph{Tear-down} phases are excluded from the measurements reported, preventing them from being tainted with extra noise from the setting-up and cleaning-up phases.
            
        \subsubsection{Measurements and Logging}
            The gathering and logging of the measurements at each period occur in two specific places: at periods' boundaries and after each execution phase.
            
            Periods' boundary measurements are taken upon the reception of the period timer-triggered signal (\emph{Measures \& log} in Figure~\ref{fig:rt-bench-logic-flowchart}). The handling of the signal prompts the taking of the measurements and their logging. Once done, a new period is started by releasing the semaphore blocking the main execution thread (i.e., \emph{the new period?} condition in Figure~\ref{fig:rt-bench-logic-flowchart}). This relationship is shown by the red dotted arrow in Figure~\ref{fig:rt-bench-logic-flowchart}. Note that this operation is only carried out if and only if the previous benchmark job has finished execution. Otherwise, the logging is filled with zeros instead, and no new jobs are released (job skipping). This also prevents logging irrelevant or misleading measurements. Deadline detection is carried out via a single boolean shared between the execution payload wrapper function and periodic signal handler. The flag is asserted when execution completes and de-asserted when logging completes.
            
        \subsubsection{Memory Footprint Watchdog}
            Upon request from the user via the provided command line options, a memory utilization watchdog is enabled through the alteration of memory allocation functions, namely, \texttt{malloc()}, \texttt{free()}, and \texttt{mmap()}. Following the framework scheme, the watchdog life-cycle is characterized by three phases: initialization, execution, and tear-down.
            
            During initialization phase, the watchdog calls the \texttt{mallopt()} function in order to pre-allocate the user-specified amount of memory and disables the \texttt{mmap()} function. A preventive allocation of the requested memory space ensures that the allocated limit is never exceeded without requiring OS modifications.
            The functions \texttt{malloc()} and \texttt{mmap()} are wrapped such that, during the execution, any call to one of these two functions results in a working set size check. In case of failure, the benchmark is terminated.
            During the tear-down phase, the watchdog is disabled, meaning \texttt{mmap()} and \texttt{mallopt()} are re-enabled and their initial parameters are restored.
            
        \subsubsection{Memory Usage Profiling}
            If requested by the user via the command line options and available on the target platform\footnote{At the time of writing, only on ARM Cortex-A53}, a thread in charge of monitoring the performance counters can be launched. In our case, the thread monitors and logs the performance counter associated with the L2 Refills.
            \newtext{To mitigate the impact on the core logic thread, it is recommended to launch it on another core (see Section~\ref{subsec:using_rtbench}) and to reduce the \emph{monitoring sample period}.}
            
            Unlike the core mechanism, the objective of this thread is to log measurements during the benchmark execution phases, instead of simply measuring before and after each execution.
            As shown in \autoref{fig:rt-bench-logic-flowchart}, the thread is launched at the initialization phase and consists of a doubly-nested loop. The first step consists of a comparison, checking whether the thread is enabled by the main thread. The change of status is operated via a shared variable asserted at the initialization phase and de-asserted at the tear-down phase. In the latter case, it leads to the benchmark's termination. Otherwise, the thread enters a ``sample-log-wait'' loop as long as the benchmark remains in its execution phase (i.e., the \emph{Bmark. running?} condition changes before and after the \emph{Execution} block).
            
            At the time of writing, the monitoring thread only samples the L2 Refills performance counter. This limitation is an implementation artifact that can be easily addressed.
            In addition, due to the close link between the performance counters and the platform implementing them, enabling support is not straightforward. The version of RT-Bench evaluated in Section~\ref{sec:eval} relies on the Linux Perf~\cite{perf} and we specifically evaluate only events specific to ARM Cortex-A53 CPUs (Table~\ref{tab:feat-comp}).
        
    \subsection{Harnessing Functions and Extension}
        \label{subsec:harnessing_functions_and_extension}
        The capability of RT-bench to enable any benchmark with the set of desired features mentioned in Section~\ref{sec:design} transparently is only possible if the benchmark has the three \emph{harnessing points}.
        
        From a practical point of view, the ``adapted'' benchmarks must implement these harnessing points. Each of them is a function with an immutable name and a clear objective.
        The \emph{Initialization} harness point is implemented as the function \texttt{benchmark\_init()}. It is in charge of initializing all the resources needed during the execution of the workload. Typically, memory allocation, variable initialization, and thread creation are carried out in the \emph{Initialization} harness point.
        The \emph{Execution} harness point is implemented as the function \texttt{benchmark\_execution()}. As the name suggests, this function consists of the workload implementation that uses the previously-set variables.
        The \emph{Tear-down} harness point is implemented as the function \texttt{benchmark\_teardown()} and is the one in charge of freeing the resources used and, if desired, posting the obtained results. In other words, it ensures a clean termination.
        
        \newtext{
        Adapting an existing benchmark to RT-Bench requires the end-user to implement the three aforementioned harnessing functions, identify the relevant code segment corresponding to each harnessing function, and move the segments in the adequate functions.
        These alterations might seem heavy; however, in reality, most benchmarks already follow a form of setup-execute-teardown organization.
        Naturally, the initial organization of the benchmark to be adapted dictates the effort required.
        As an indication, we report on the changes and efforts required to adapt SD-VBS's \texttt{disparity} and \texttt{pca} via their number of changed lines of code using the cloc~\cite{cloc} tool.
        The \texttt{disparity} benchmark modifications amounts to 19 modified, 17 added and 12 removed SLOCs (or source lines of code) whereas modifications to \texttt{pca} amounts to 68 modified, 182 added and 216 removed SLOCs. 
        Overall, the adaptation of the ten benchmarks composing the SD-VBS suite required a total of 302 modified, 925 added, and 689 removed SLOCs.
        }
        
        \newtext{
        The end-user is free to define the content of the harnessing functions as desired. Nonetheless, when multi-threading is required, we recommend implementing the workload using the fork-join approach in each relevant function. In other words, we recommend that every thread created within a harnessing function is destroyed within the same function.
        }
        
        
        \newtext{
        On the compilation side, an executable of the ported benchmark can be obtained via mainstream tools such as \texttt{gcc}. The ported source files shall not implement an entry function (i.e., \texttt{main}). Instead, the RT-Bench core interface must be linked in. As such, additional file directories must be added to the include path. This translates in utilizing \texttt{gcc}'s \texttt{-I/path/to/rt-bench/base/} option in addition to any benchmark-specific compilation flags and options.
        }
        
    \subsection{Common Input Interface}
        \label{subsec:using_rtbench}
        As per the design goals presented in Section~\ref{sec:design}, any benchmark yielded by the RT-Benchmark Generator benefits from the same set of features, composing the homogenized input interface. Each of these features can be tailored via the enabled command-line options. The options under only represent a subset of the options made available to all benchmarks by RT-bench:
        \begin{itemize}
        	\item[\texttt{-p}]: Relative period of a single benchmark execution;
        	\item[\texttt{-d}]: Relative deadline of a single benchmark execution;
        	\item[\texttt{-l}]: Log level;
        	\item[\texttt{-c}]: Core affinity;
        	\item[\texttt{-f}]: FIFO scheduling with specified priority;
        	\item[\texttt{-m}]: Memory limit;
        	\item[\texttt{-t}]: Number of tasks to execute before termination;
        	\item[\texttt{-b}]: Benchmark specific arguments and options;
        	\item[\texttt{-P}]: SCHED\_DEADLINE period;
        	\item[\texttt{-D}]: SCHED\_DEADLINE deadline;
        	\item[\texttt{-T}]: SCHED\_DEADLINE runtime;
        	\item[\texttt{-M}]: Enable PMC monitoring thread;
        	\item[\texttt{-C}]: Core affinity for PMC monitoring thread;
        	\item[\texttt{-B}]: PMC monitoring sample period;
        \end{itemize}
        The options listed above constitute the main options used in the Evaluation (see Section~\ref{sec:eval}). An exhaustive list of the options, together with additional details, is provided in the project documentation.
        
    \subsection{High-level Automated Tests}
        \label{subsec:high-level-automated-tests}
        The provided base scripts written in \texttt{Python3} constitute a collection of utility functions implementing profiling and real-time minded experiments. These tests have been used in the article's evaluation section (Section~\ref{sec:eval}) to highlight the capability of the RT-Bench framework. At the time of writing, the set includes six experiments:
        
        \vspace*{0.1cm}
        \pgrph{Minimum WSS.}{This test aims at \newtext{empirically deriving} the least amount of memory footprint required by the benchmark. To do so, the test explores \newtext{the memory size allocation space via a binary search using the fact that the \emph{memory watchdog} terminates any benchmark exceeding the user-defined allocation limited as the discriminant. I.e., if the program is terminated by the watchdog, the WSS is larger than the imposed limit; conversely, if the benchmark completed correctly, the WSS is smaller than (or equal to) the considered amount.}}
        
        \vspace*{0.1cm}
        \pgrph{WCET.}{Thanks to the metrics reported by the RT-benchmarks, determining a benchmark's WCET can be empirically obtained via subsequent execution batches. The provided test \emph{explores} candidate values by setting a default deadline value and \emph{validates} by using the maximal observed execution time as the deadline for the following batch. The WCET is set as the minimum deadline value that reliably \newtext{prevents} misses when the benchmark is executed in isolation.}
        
        \vspace*{0.1cm}
        \pgrph{Schedulability Test.}{Based on previously established WCETs, this test looks at the rate of schedulable/unschedulable jobs as a function of the task's utilization. In this case, the deadlines are determined by dividing the previously-derived WCET by the target utilization. The test starts at $5\%$ utilization and goes up to $100\%$ in steps of $5\%$.}
        
        \vspace*{0.1cm}
        \pgrph{Caches Miss Rate.}{Leveraging the performance counters reported in the output interface, the Cache Miss Rate metric can be easily obtained by computing the ratio between the cache references and cache refills events. The test is applied on any available cache level.}
        
        \vspace*{0.1cm}
        \pgrph{Memory and CPU Intensity.}{This test investigates if a benchmark is CPU- or memory-bound by inspecting the ratio between the L2 cache misses and the number of retired instructions, two metrics natively reported by RT-Bench.}
        
        \vspace*{0.1cm}
        \pgrph{Memory Usage Profiling.}{Perhaps more importantly than knowing whether a benchmark is memory-bound, understanding the run-time demand is crucial for any system under memory bandwidth regulation. This test highlights the memory consumption phases a benchmark displays.}
        
        \vspace*{0.1cm}
        In any of the aforementioned tests, basic manipulations are performed. For instance, (1) before the benchmark execution, all tasks on the device are migrated to one core (often core 0) if requested by the user (2) measurements obtained are automatically plotted, and (3) co-running interference tasks are launched on other cores.
\section{Evaluation}
    \label{sec:eval}

    \begin{table}[]
        \caption{Comparative table of the evaluation platforms}
        \vspace{-0.3cm}
        \label{tab:evaluation_platforms}
        \begin{tabular}{ccc}
            \toprule
                         & Xilinx ZCU102                            & AMD RYZEN 9 5900HS        \\
            \midrule
            ISA          & ARM64                                     & x86\_64                   \\
            CPU          & 4$\times$Cortex-A53 (@1.5GHz)             & 8$\times$CPU (@3-4.6GHz)  \\
            L1           & 32KB+32KB I \& D caches                   & 8KB+8KB I \& D caches     \\
            L2           & 1MB Unified cache                         & 4MB Unified cache         \\
            L3           & \textbf{-}                                & 16MB Unified cache        \\
            DRAM         & 4GB DDR4                                  & 32GB DDR4                 \\
            Linux        & 5.4.14                                    & 5.16.9                    \\
            GCC          & 9.4.0                                     & 10.3                      \\
            \bottomrule
        \end{tabular}
        \vspace{-0.3cm}
    \end{table}

    This section showcases the capabilities and user-friendliness of the proposed framework, RT-Bench. The evaluation presented in this section consists in the set of experiments listed in Section~\ref{subsec:high-level-automated-tests}: finding the minimum WSS (Section~\ref{subsec:eval-wss}), determining the observed WCET (Section~\ref{subsec:eval-wcet}), performing the schedulability test (Section~\ref{subsec:eval-sched}), studying the cache miss rates (Section~\ref{subsec:eval-caches-miss-rate}), understanding whether the benchmark is memory or CPU bound (Section~\ref{subsec:eval-mem-cpu-intensity}), and observing the evolution of the memory consumption at run-time (Section~\ref{subsec:eval-mem-usage-profile}).
    
    The experiments have been performed on two different platforms: the Xilinx ZCU102 development board and the widely available AMD RYZEN 9 CPU model. Their architecture specifications and the version of the software tools (e.g., Linux kernel version and GCC version) are displayed in Table~\ref{tab:evaluation_platforms}. From now on, the Xilinx ZCU102 is referred to as the ``ARM platform'' whereas the AMD RYZEN 9 is referred to as the ``x86 platform''.

    Throughout the evaluation, the RT-Bench's capabilities are shown by using benchmarks issued from a RT-Bench adapted version of the San Diego Vision Suite (or SD-VBS)~\cite{SD-VBS}. The exact benchmarks considered are \texttt{disparity}, \texttt{mser}, \texttt{localization}, \texttt{tracking}, and \texttt{sift}. In addition, all the available input sizes but \texttt{test}, \texttt{qcif}, and \texttt{full\_hd} have been considered, that is \texttt{sim\_fast}, \texttt{sim}, \texttt{sqcif}, \texttt{cif}, and \texttt{vga} (ordered by increasing size). For tests requiring an interfering co-runner, instances of the ``bandwidth'' benchmark issued from a RT-Bench adapted version of IsolBench~\cite{IsolBench} are launched. The interfering task instances sweep across a dedicated 100MB-wide buffer. Their number and their memory access mode (i.e., read or write) depend on the test performed and the platform capabilities.
    
    In each experiment presented in this section, the benchmark under analysis is run using \texttt{SCHED\_FIFO} and is assigned a priority of $99$. Likewise, interfering co-runners are assigned a priority of $99$. The \emph{RT-Throttling} is turned off, allowing for a $100\%$ CPU utilization.

    \subsection{Minimum Working Set Size Test}
        \label{subsec:eval-wss}
\begin{figure}
        	\centering
        	\includegraphics[width=0.85\columnwidth]{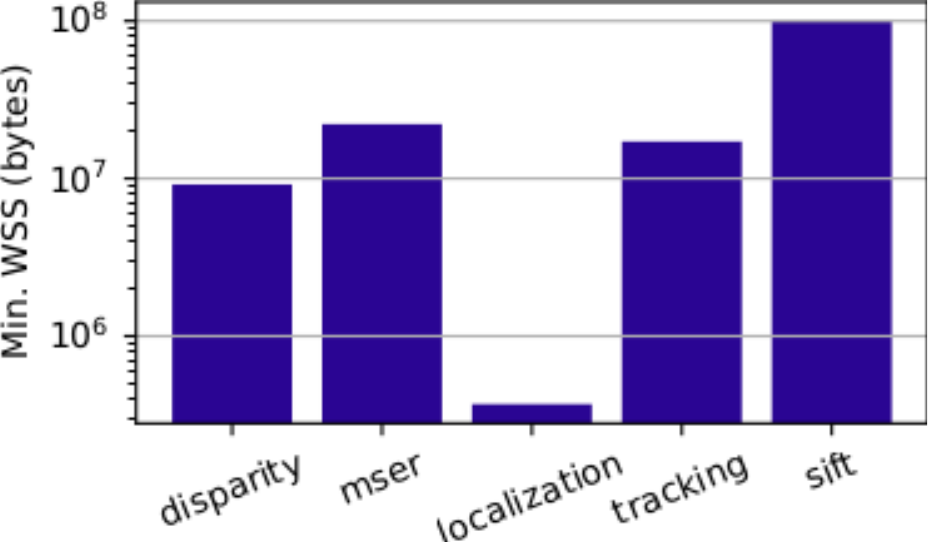}
        	\Description{A bar graph containing minimum working set sizes for the vga
        		input size of different SD-VBS benchmarks.}
        	\caption{SD-VBS benchmarks minimum WSS for \texttt{vga} input.}
        	\label{fig:vga-wss}
        \end{figure}

\begin{figure}
        	\centering
        	\includegraphics[width=0.85\columnwidth]{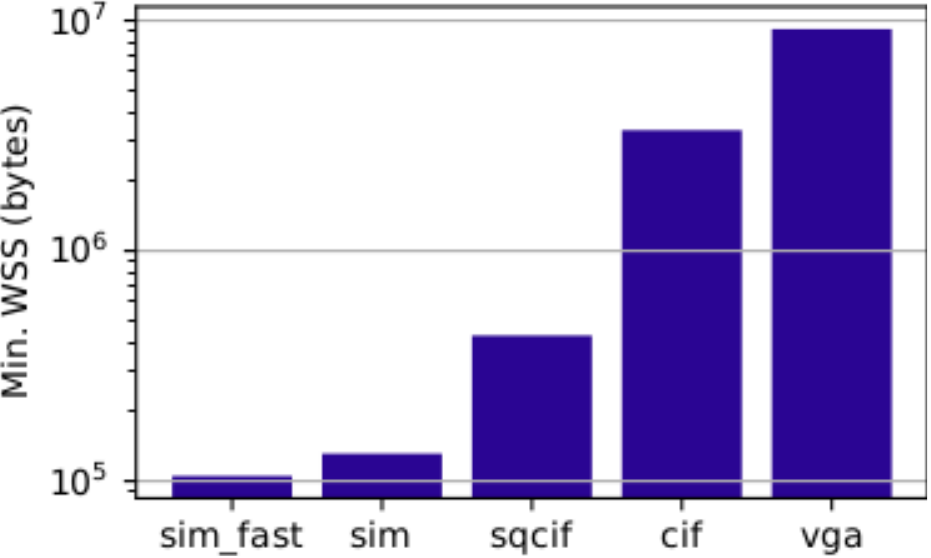}
        	\Description{A bar graph containing minimum working set sizes for different input sizes of the SD-VBS disparity benchmark.}
        	\caption{\texttt{Disparity}'s minimum WSS for different inputs.}
        	\label{fig:disparity-wss}
        \end{figure}

        First, this experiment investigates the WSS of the supported SD-VBS benchmarks (\autoref{fig:vga-wss}). Next, we place our emphasis on the WSS of \texttt{disparity} for all the available inputs (\autoref{fig:disparity-wss}).
        In both \autoref{fig:vga-wss} and \ref{fig:disparity-wss} the minimal WSS found is reported by the height of the bars (y-axis in log scale).
        This set of experiments is only carried out on the x86 platform due to space constraints.

        \autoref{fig:vga-wss} shows that, for the \texttt{vga} input, all the benchmarks require at least 10MB of main memory. Only \texttt{sift} and \texttt{localization} do not follow the rule as the former requires 100MB and the latter requires 1MB. However, as highlighted by \autoref{fig:disparity-wss}, the minimum required memory footprint is dependent on the input. In fact, one can observe that the WSS for a \texttt{vga} input is orders of magnitude bigger than that for a \texttt{sim\_fast} input. Note that the observed size order matches the input size order.
        
    \subsection{Worst case execution time test}
        \label{subsec:eval-wcet}
        In this experiment, the WCET test is used to understand the intrinsic behavior of the benchmarks when running in isolation (i.e., alone) and when they face memory interference from other cores. We present tests run on both the x86 and the ARM platforms.
        
        To represent the distribution of the measured execution times, a violin plot was chosen.
        Each violin is associated with a benchmark running a \texttt{vga} input on the $x$-axis, and the $y$-axis reports their measured execution time in seconds.
        Each violin is composed of three horizontal lines representing the minimum, maximum and average measurements. The width of the violins represents the distributions of all the measurements.

        \begin{figure}
        	\centering
        	\includegraphics[width=0.85\columnwidth]{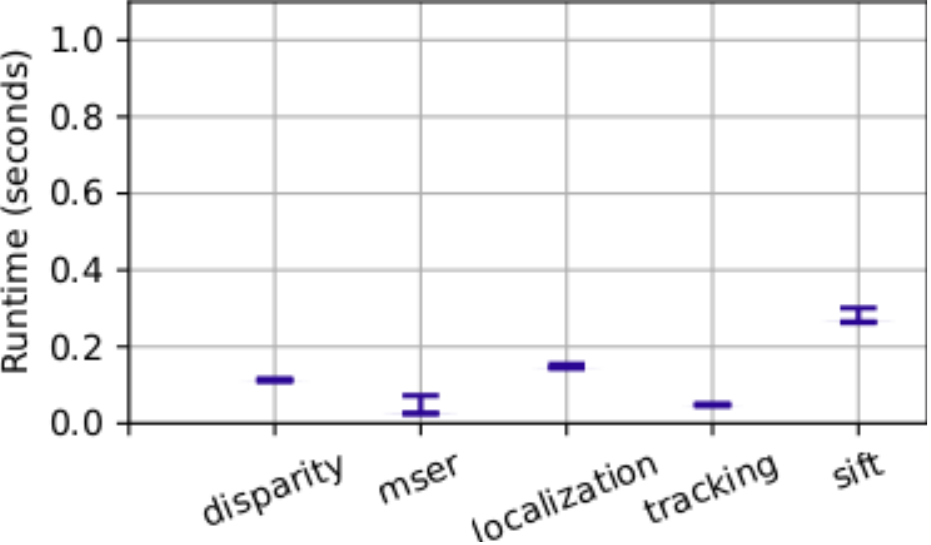}
        	\Description{A violin graph containing the worst execution times for the vga
        		input size of different SD-VBS benchmarks.}
        	\caption{SD-VBS benchmarks WCET on x86 with \texttt{vga} input.}
        	\label{fig:vga-wcet}
        \end{figure}
        
        \begin{figure*}
            \centering
            \begin{subfigure}[b]{0.49\textwidth}
                \centering
            	\includegraphics[width=0.9\columnwidth]{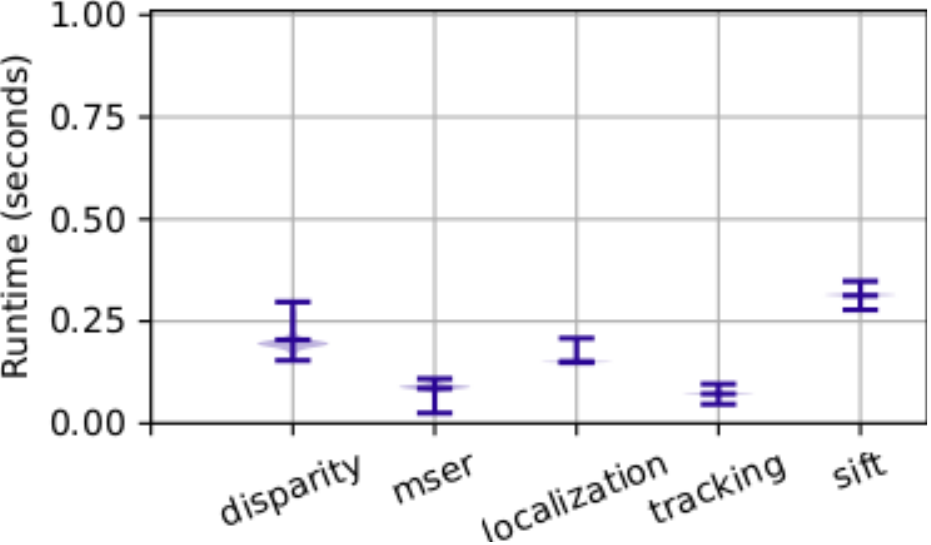}
            	\Description{A violinplot containing WCET tests for the vga input
            		size of different SD-VBS benchmarks.}
            	\caption{4 cores interference.}
            	\label{fig:vga-wcet-4}
            \end{subfigure}
            \hfill
            \begin{subfigure}[b]{0.49\textwidth}
                \centering
            	\includegraphics[width=0.9\columnwidth]{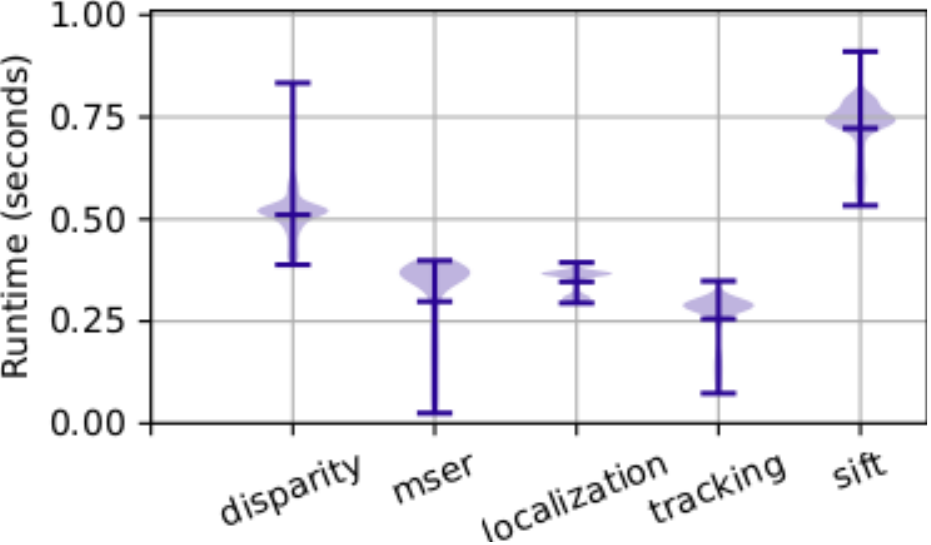}
            	\Description{A violinplot containing WCET tests for the vga input
            		size of different SD-VBS benchmarks.}
            	\caption{6 cores interference.}
            	\label{fig:vga-wcet-6}
            \end{subfigure}
            \caption{SD-VBS benchmarks WCET tests on x86\_64 on \texttt{vga} input with interference.}
            \label{fig:vga-wcet-interf}
        \end{figure*}
        
        On the x86 platform, three scenarios are explored: (1) WCET in isolation (\autoref{fig:vga-wcet}), (2) WCET with 2 read and 2 write interfering cores (\autoref{fig:vga-wcet-4}), and (3) WCET with 6 write interfering cores (\autoref{fig:vga-wcet-6}).
        \autoref{fig:vga-wcet} shows that without interfering processes, most of the benchmark execution samples do not present high variance. Conversely, \texttt{mser} and \texttt{sift} are the most likely to suffer from inter-core interference.
        This intuition is confirmed by \autoref{fig:vga-wcet-4} which shows that, under interference, all benchmarks see their execution time distributions being stretched. 
        In contrasts, \autoref{fig:vga-wcet-6} shows that memory interference created by 6 cores writing data introduces higher variations in execution times. \texttt{Disparity} and \texttt{sift} are the most impacted with their WCET increased by twofold in \autoref{fig:vga-wcet-4}, and eight-fold and threefold in \autoref{fig:vga-wcet-6}, respectively.
        
        \begin{figure*}
            \centering
            \begin{subfigure}[b]{0.49\textwidth}
                \centering
            	\includegraphics[width=0.90\columnwidth]{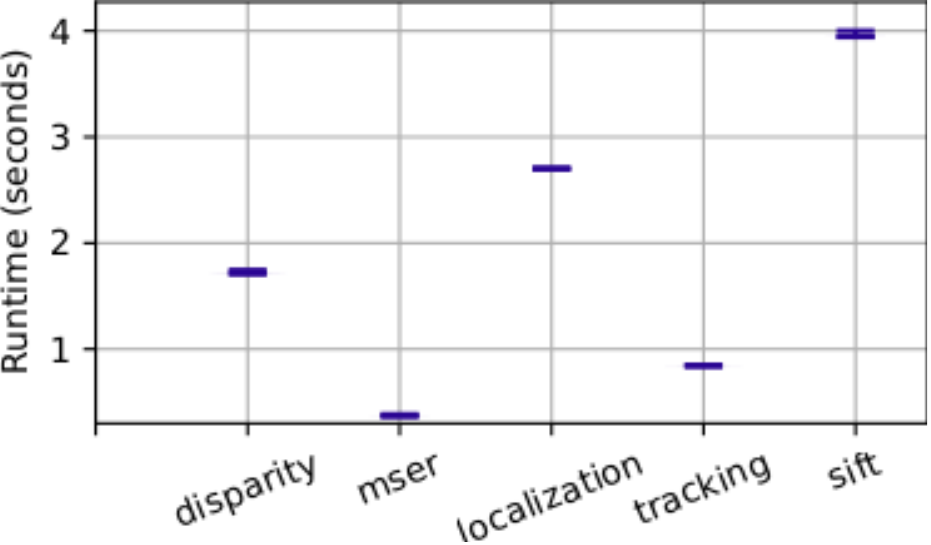}
            	\Description{A violinplot containing WCET tests for the vga input
            		size of different SD-VBS benchmarks.}
            	\caption{no interference.}
            	\label{fig:zcu-wcet-no-interf}
            \end{subfigure}
            \hfill
            \begin{subfigure}[b]{0.49\textwidth}
                \centering
            	\includegraphics[width=0.90\columnwidth]{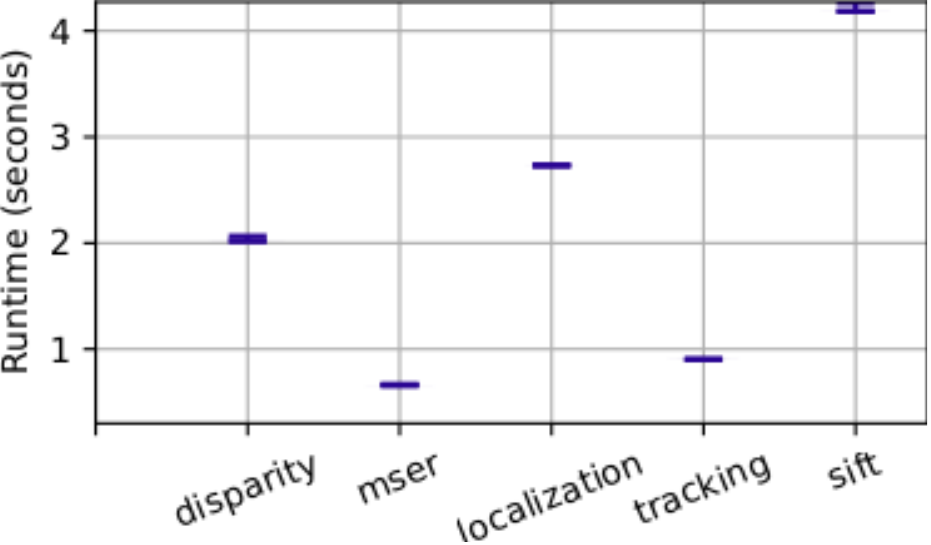}
            	\Description{A violinplot containing WCET tests for the vga input
            		size of different SD-VBS benchmarks.}
            	\caption{2 cores interference.}
            	\label{fig:zcu-wcet-interf}
            \end{subfigure}
            \caption{SD-VBS benchmarks WCET tests on ARM64 with \texttt{vga} input.}
            \label{fig:zcu-wcet}
        \end{figure*}
        
        On the ARM platform, two similar scenarios have been explored: WCET in isolation \autoref{fig:zcu-wcet-no-interf} and WCET with 2 write-interfering cores \autoref{fig:zcu-wcet-interf}.
        Unlike the x86 platform, the effect of interference creates a more consistent execution time distributions and only leads to longer execution times.
        However, as with the x86 scenarios, \autoref{fig:zcu-wcet-no-interf} and \ref{fig:zcu-wcet-interf} show that \texttt{disparity} and \texttt{sift} are the most impacted by interference.

    \subsection{\newtext{Deadline Miss Ratio Test}}
        \label{subsec:eval-sched}
\begin{figure*}
            \centering
            \begin{subfigure}[b]{0.49\textwidth}
                \centering
            	\includegraphics[width=\columnwidth]{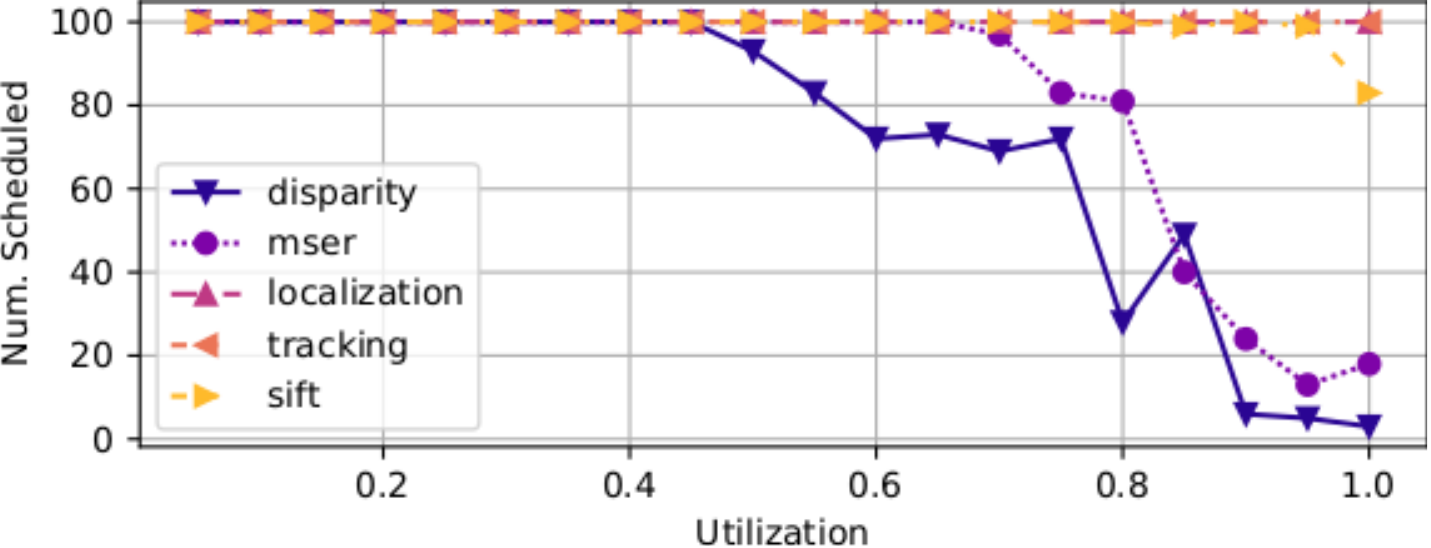}
            	\Description{A bar graph containing schedulability tests for the vga input
            		size of different SD-VBS benchmarks.}
            	\caption{4 cores interference.}
            	\label{fig:vga-sched-4}
            \end{subfigure}
            \hfill
            \begin{subfigure}[b]{0.49\textwidth}
                \centering
            	\includegraphics[width=\columnwidth]{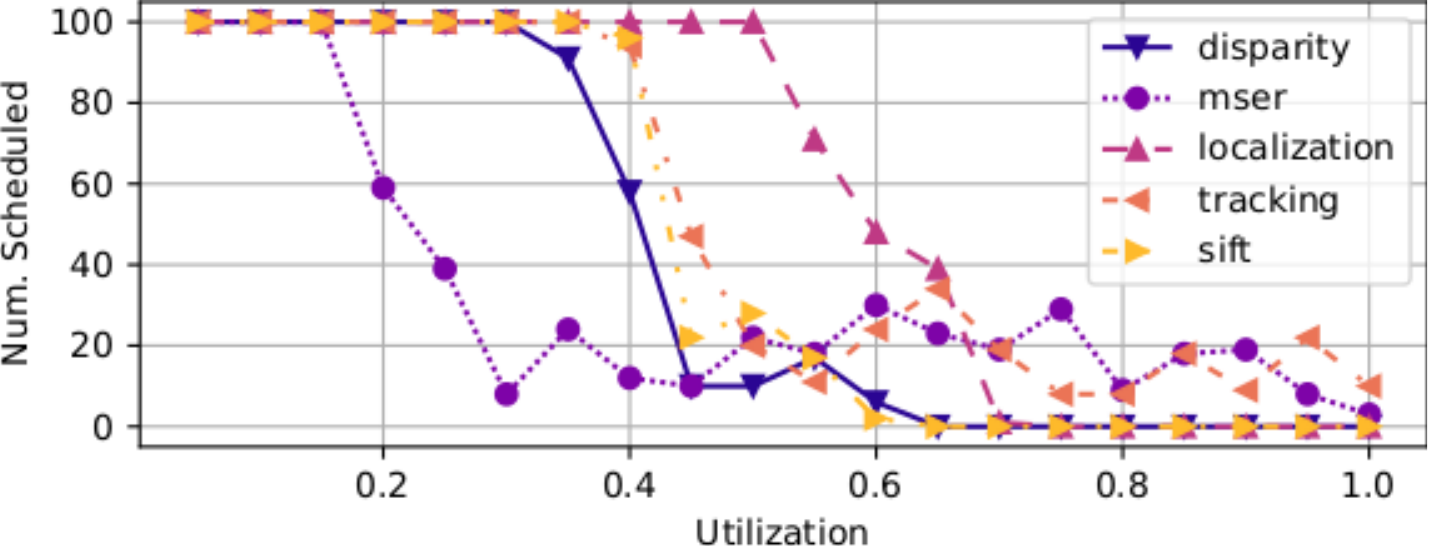}
            	\Description{A bar graph containing schedulability tests for the vga input
            		size of different SD-VBS benchmarks.}
            	\caption{6 cores interference.}
            	\label{fig:vga-sched-6}
            \end{subfigure}
            \caption{SD-VBS benchmarks schedulability tests on x86\_64 on \texttt{vga} input with interference.}
            \label{fig:vga-sched}
        \end{figure*}

\begin{figure}
        	\centering
        	\includegraphics[width=\columnwidth]{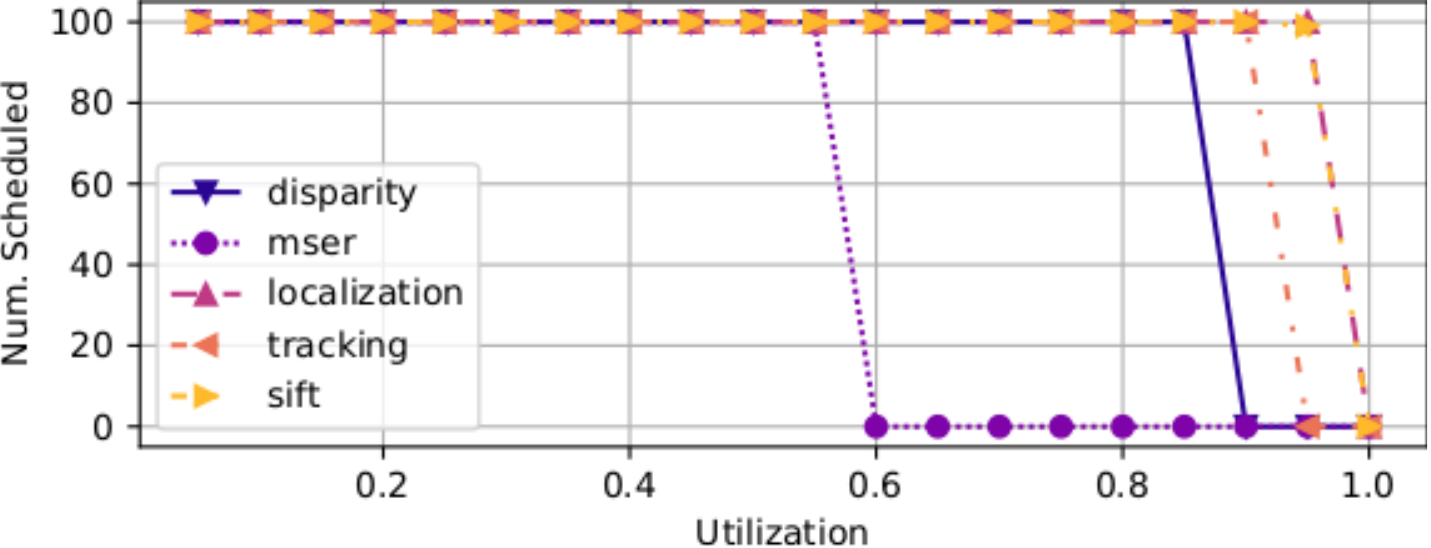}
        	\Description{A plot containing schedulability tests for the vga input of different SD-VBS disparity benchmark.}
        	\caption{SD-VBS disparity schedulability test on ARM64 with \texttt{vga} input and 2 cores that produce interference.}
        	\label{fig:zcu-sched}
        \end{figure}

        To gain insight into the schedulability of the chosen benchmarks at a certain system load, two scenarios on the x86 platform and one scenario on the ARM platform are shown. 
        On the x86 platform, \autoref{fig:vga-sched-4} shows the effect of two read and two write interfering cores, while \autoref{fig:vga-sched-6} shows the effect of six write interfering cores.
        On the ARM platform, there is only one scenario with two writing cores that generate interference, as shown by \autoref{fig:zcu-sched}.
        In both \autoref{fig:vga-sched} and \autoref{fig:zcu-sched}, the x-axis of the figures shows the utilization value, while the y-axis shows the number of benchmarks that met the deadline.
        
        \autoref{fig:vga-sched-4} shows that only \texttt{mser} and \texttt{disparity} are severely impacted by the interference on the other four cores. While the impact on the other benchmarks is minimal.
        However, changing the interference pattern to six cores will severely impact all the benchmarks, keeping \texttt{mser} and \texttt{disparity} as the most impacted ones, as \autoref{fig:vga-sched-6} shows.
        
        As \autoref{fig:zcu-sched} shows, the ARM platform has a more predictable behavior than the x86 platform, having all the benchmarks meet the deadline or failing when the deadline gets too short to allow the benchmark to complete the execution with two writing cores that produce interference. As on the x86 scenarios, the most impacted benchmarks are \texttt{mser} and \texttt{disparity}.
        
    \subsection{Caches Miss Rate}
        \label{subsec:eval-caches-miss-rate}
\begin{figure}
            \centering
            \includegraphics[width=0.80\columnwidth]{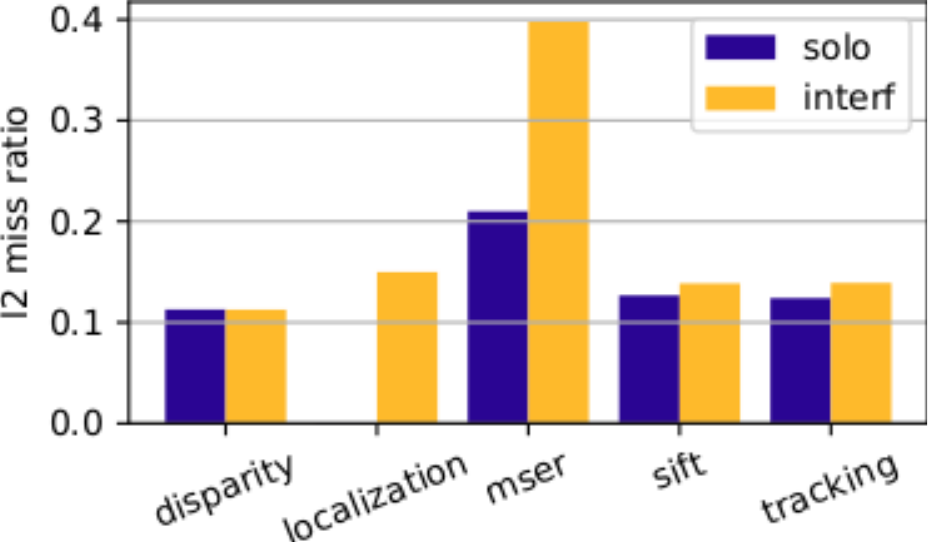}
            \Description{L2 cache miss rate for each the chosen SD-VBS benchmarks with and without interference.}
            \caption{SD-VBS benchmarks' L2 cache miss-rate with and without interference (\texttt{vga} input).}
            \label{fig:zcu-cache}
        \end{figure}

        The cache miss rate experienced by a benchmark is a widely used metric to show how reliant on memory a benchmark is and the extent of memory interference impact. The L2 miss-rate experienced by the benchmarks running on the ARM platform is shown in \autoref{fig:zcu-cache} (the bar clusters). In each bar cluster, the miss rate when running in isolation is drawn in blue (referred to as ``solo''), whereas the observed miss rate under a two cores write contention is drawn in yellow (referred to as ``interf'').
        \autoref{fig:zcu-cache} highlights the existence of two categories. On the one hand, \texttt{disparity}, \texttt{sift}, and \texttt{tracking} are marginally impacted, hinting at a low temporal data locality (if the data is not reused later on, it does not matter whether it is evicted by an interfering task).
        On the other hand, \texttt{localization} and \texttt{mser} display a higher sensitivity to memory interference, hinting at a high temporal data locality. Remarkably, \texttt{mser} constitutes a hybrid case as it naturally displays a high miss rate in isolation and high sensitivity to memory interference.
        
    \subsection{Memory and CPU Intensity}
        \label{subsec:eval-mem-cpu-intensity}
\begin{figure}
        	\centering
        	\includegraphics[width=0.80\columnwidth]{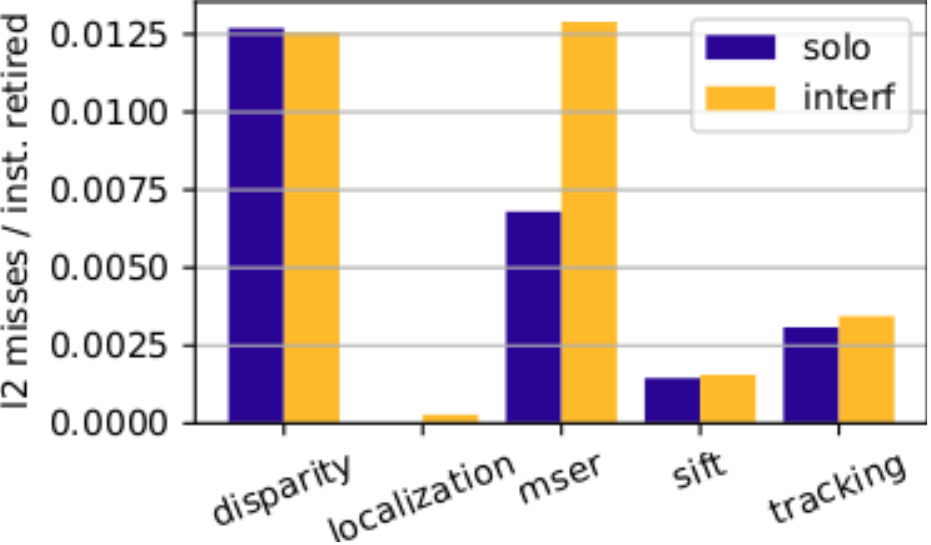}
        	\Description{A bar graph with the ratio between L2 cache misses and instructions retired for each the chosen SD-VBS benchmarks with and without interference.}
        	\caption{SD-VBS benchmarks' L2 cache miss-rate over instruction retired ratio with and without interference (\texttt{vga} input).}
        	\label{fig:zcu-perf}
        \end{figure}

        To get further insight into the execution behavior of the chosen benchmarks, their ratio between the cache misses, and the number of instructions retired can be analyzed.
        This analysis is portrayed for the ARM platform by \autoref{fig:zcu-perf}, which shows, in a bar graph, the aforementioned ratio for all the benchmarks with and without interference.
        
        From \autoref{fig:zcu-perf} it can be deducted if a benchmark is more memory or CPU bound. \texttt{localization} is an example of a CPU bound benchmark, while \texttt{disparity} is an example of a memory-bound benchmark. As for the previous test, the benchmark most impacted by interference is \texttt{mser}.
        Comparing \autoref{fig:zcu-cache} and \autoref{fig:zcu-perf} insight on how the cache misses affect the benchmark execution can be gained. While \texttt{sift} and \texttt{tracking} have more or less the same amount of cache misses, \texttt{sift} is more CPU bound than \texttt{tracking}, due to a smaller ratio between cache misses and instructions retired. It can also be inferred that \texttt{localization} is the most CPU-bound benchmark, since it has the lowest ratio between cache misses and instructions retired.  
        
    \subsection{Memory Usage Profile}
        \label{subsec:eval-mem-usage-profile}
\begin{figure}
            \begin{subfigure}[b]{0.49\textwidth}
            	\centering
            	\includegraphics[width=0.85\columnwidth]{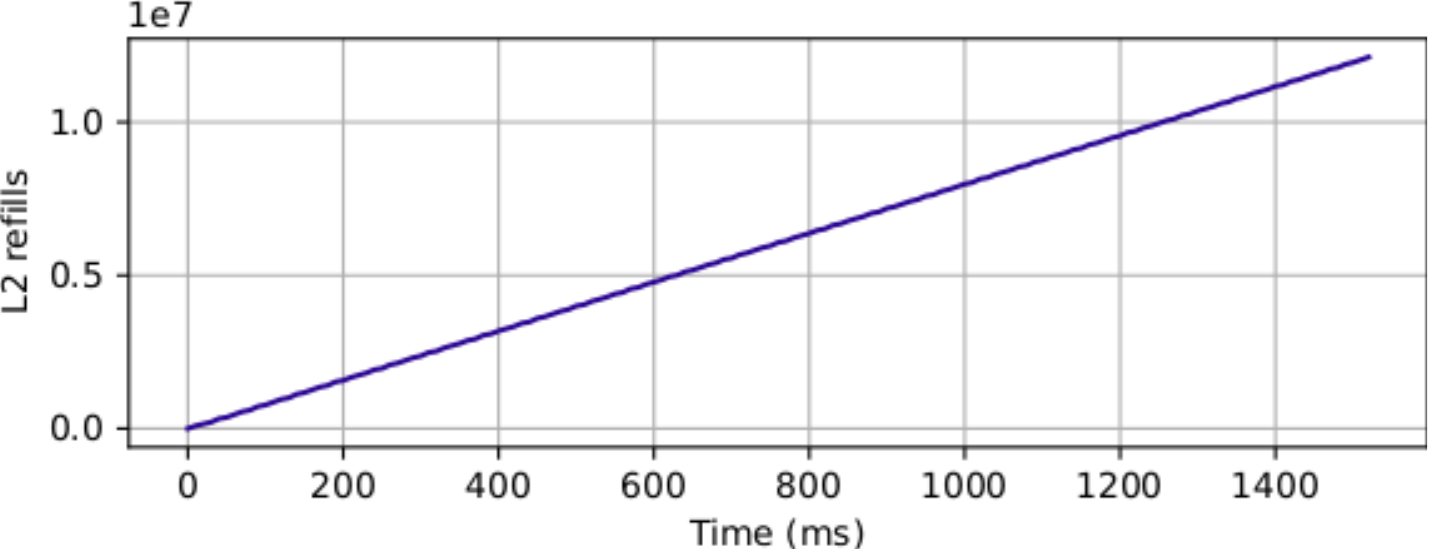}
            	\Description{A plot with the L2 cache misses for SD-VBS disparity taken as a snapshot every 10 ms.}
            	\caption{SD-VBS \texttt{disparity}.}
            	\label{fig:disparity-zcu-perf-monitor}
            \end{subfigure}
            \hfill
            \begin{subfigure}[b]{0.49\textwidth}
            	\centering
            	\includegraphics[width=0.85\columnwidth]{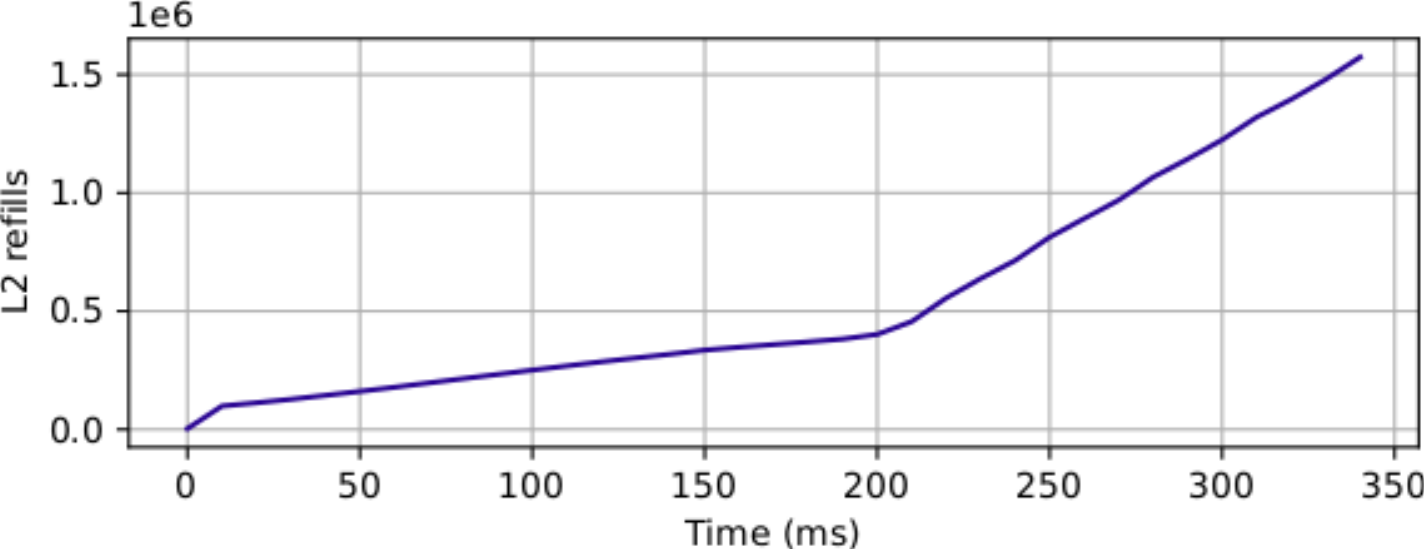}
            	\Description{A plot with the L2 cache misses for SD-VBS mser taken as a snapshot every 10 ms.}
            	\caption{SD-VBS \texttt{mser}.}
            	\label{fig:mser-zcu-perf-monitor}
            \end{subfigure}
            \hfill
            \begin{subfigure}[b]{0.49\textwidth}
            	\centering
            	\includegraphics[width=0.85\columnwidth]{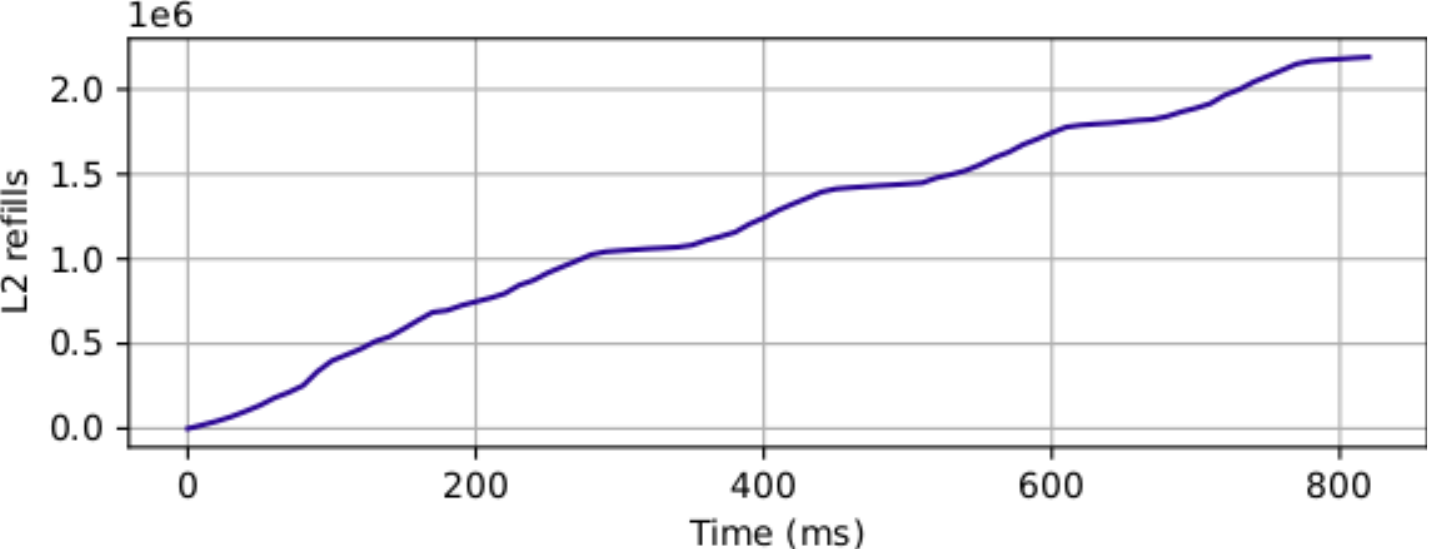}
            	\Description{A plot with the L2 cache misses for SD-VBS tracking taken as a snapshot every 10 ms.}
            	\caption{SD-VBS \texttt{tracking}.}
            	\label{fig:tracking-zcu-perf-monitor}
            \end{subfigure}
        	\caption{The plot shows the total L2 cache refills during a single benchmark execution using the \texttt{vga} input.}
        	\label{fig:zcu-perf-monitor}
        \end{figure}

        A memory usage profile can help identify how a benchmark uses memory during its execution. \autoref{fig:zcu-perf-monitor} show the memory profiles of \texttt{disparity}, \texttt{mser} and \texttt{tracking} during their execution with a plot of the L2 cache misses on the y-axis and the time on the x-axis.
        
        \autoref{fig:disparity-zcu-perf-monitor} shows that \texttt{disparity} is always accessing memory during its execution; this explains why it is so impacted by interference, even if it does not have a significant increase in cache misses when interfering processes are present. 
        Instead, \texttt{mser} has three well-defined phases in which its memory access pattern changes significantly, explaining why it is impacted so heavily by interference in \autoref{fig:zcu-perf}. \texttt{tracking} has different phases with a different memory access pattern. This pattern highlights that it benefits from temporal and spatial locality in the data.

        \subsection{\newtext{RT-Bench Framework Overhead}}
        \newtext{To quantify the overhead of the framework, an ad-hoc benchmark, \texttt{overhead}, has been used. The main payload of the \texttt{overhead} benchmark consists of an empty function. Therefore, the benchmark execution time will be dominated by the time spent to execute RT-Bench's core logic.}
        \begin{table}[]
            \caption{Rt-Bench overhead measurements}
            \label{tab:overhead}
            \centering
            \begin{tabular}{lcccc}
                \toprule
                Platform & Min & Mean & Max & Std \\
                \midrule
                 x86 ($\mu s$) & $5.49$ & $24.57$ & $515.99$ & $13.99$  \\
                 ARM no Perf ($\mu s$) & $1.36$ & $30.63$ & $95.91$ & $8.39$  \\
                 ARM Perf ($\mu s$) & $25.33$ & $44.35$ & $115.90$ & $6.06$  \\
                 x86 (clock cycles) & $18711$ & $84987$ & $1700193$ & $47339$  \\
                 ARM no Perf (clock cycles) & $2877$ & $3086$ & $9570$ & $819$  \\
                 ARM Perf (clock cycles) & $3817$ & $4438$ & $11588$ & $605$  \\
                 \bottomrule
            \end{tabular}
        \end{table}
        \newtext{\autoref{tab:overhead} shows the measured overhead on the two platforms with and without using the performance counters. It can be observed that, on average, the overhead is contained between $20$ to $50$ $\mu s$.}
        
\section{Conclusion}
    The article presents RT-Bench, an open-source framework that aims to ease the tedious task of profiling and monitoring commonly used benchmark suites by providing a unified interface that can be built upon and re-used by the community. RT-Bench lays the foundation for a coherent benchmarking and profiling system for the real-time community. We provided an in-depth description of RT-Bench capability and outlined the main implemented features for a clean and reusable interface.
    
    Through the evaluation of RT-Bench presented in Section~\ref{sec:eval} using well-known benchmarks suites such as SD-VBS and IsolBench, we showcase how the proposed implementation drastically simplifies the gathering and post-processing of experimental data.
    
    While enabling the end-user with an interesting range of features, the presented version of RT-Bench is in its early days with a sizeable potential for community-fueled contributions and improvements. These include increasing the range of collected data, adding more performance counters, extending the provided benchmarks (including other popular suites like MiBench and TACLeBench), and extending the inputs to enable broader insight into the benchmark behavior with different inputs of the same size. Other possible avenues are the support for DAG tasks and a broader range of architectures, such as PowerPC and RISC-V. Finally, integration with IPC systems could be pursued to analyze inter-task dependencies.

\begin{acks}
The material presented in this paper is based upon work supported by the National Science Foundation (NSF) under grant number CCF-2008799. Any opinions, findings, and conclusions or recommendations expressed in this publication are those of the authors and do not necessarily reflect the views of the NSF.
Andrea Bastoni and Denis Hoornaert were supported by the Chair for Cyber-Physical Systems in Production Engineering at TUM and the Alexander von Humboldt Foundation.
\end{acks}

\bibliography{references.bib}

\end{document}